\documentclass[aps,preprint,,showpacs,floats,epsf,epsfig,nofootinbib,12pt]{revtex4}
\usepackage{graphicx}
\usepackage{epsfig}
\usepackage[dvips]{color}

\def\beq{\begin{eqnarray}}
\def\eeq{\end{eqnarray}}
\def\non{\nonumber}

\def\nslash{\rlap{\hspace{0.01cm}/}{n}}
\def\nbarslash{\rlap{\hspace{0.01cm}/}{\bar n}}
\def\vslash{\rlap{\hspace{-0.02cm}/}{v}}
\def\la{\langle}
\def\ra{\rangle}
\def\Mb{M_{\Lambda_b}}
\def\Mc{M_{\Lambda_c}}

\begin{document}
\title{\bf\Large Evaluating decay Rates and Asymmetries of $\Lambda_b$ into Light
  Baryons in LFQM}

\vspace{1cm}

\author{
 Zheng-Tao Wei$^1$,
 Hong-Wei Ke$^2$\footnote{Corresponding author, email: khw020056@hotmail.com} and
 Xue-Qian Li$^1$}

\vspace{1cm}

\address{1. School of Physics, Nankai University, Tianjin, 300071,
P.R. China,\\
2.  School of Sciences, Tianjin University, Tianjin, 300072, P.R.
China}

\vspace{1cm}

\begin{abstract}

In this work we calculate the branching ratios of semi-leptonic and
non-leptonic decays of $\Lambda_b$ into light baryons ($p$ and
$\Lambda$), as well as the measurable asymmetries which appear in
the processes, in the light front quark model (LFQM). In the
calculation, we adopt the diquark picture and discuss the
justifiability of applying the picture in our case. Our result on
the branching ratio of $\Lambda_b\to\Lambda+J/\psi$ is in good
agreement with data.  More predictions are made in the same model
and the results will be tested in the future experiments which will
be conducted at LHCb and even ILC.

\end{abstract}
 \maketitle
\vspace{1cm}

\section{introduction}
As well known, the $\Lambda_b$ weak decay may give us abundant
information about CKM elements, so that it stands as a complementary
field to the meson decays. These processes are also good probes for
the factorization hypothesis which has been extensively explored for
dealing with hadronic transitions \cite{mix,Factorization}. Recently
many semi-leptonic and non-leptonic decays of $\Lambda_b$ are
observed and measured \cite{DELPHI,PDG06}, moreover the LHCb is
expected to accumulate a large data-sample of b-hadrons to offer a
unique opportunity for studying $\Lambda_b$, thus we would like to
investigate the $\Lambda_b$ weak decay more systematically. As for
the $\Lambda_b$ decays  the key is how to evaluate the form factors
which parameterize the hadronic matrix elements. There are many
approaches advocated to this aspect \cite{AHN}. In our previous
paper \cite{befor} we studied $\Lambda_b$ to $\Lambda_c$ weak decay
in the light-front quark model \cite{light} and the results seem to
be quite reasonable.

The light-front quark model is a relativistic quark model based on
the light-front QCD \cite{light}. The basic ingredient is the
hadron light-front wave function which is explicitly
Lorentz-invariant. The hadron spin is constructed using the Melosh
rotation. The light-front approach has been widely applied to
calculate various decay constants and form factors for the meson
cases \cite{meson1,meson2,CCH1,CCH2,HW}.

In our earlier work, we adopted the diquark picture for baryons
\cite{befor} which especially is well explored and proved to be a
good approximation for such processes where the diquarks are not
broken during the transition. Indeed, it has been known for a long
time that two quarks in a color-antitriplet state attract each other
and may form a correlated diquark \cite{DJS}. The diquark picture of
baryons is considered to be appropriate for low momentum-transfer
processes \cite{kroll,wilczek,yu,MQS}. Concretely, under the diquark
approximation, $\Lambda_b$ and $\Lambda_c$ are of the
one-heavy-quark-one-light-diquark(ud) structure which is analogous
to the meson case.

In this paper we will apply these method to $\Lambda_b$ decaying
into light hadrons such as proton or $\Lambda$ which is made of
three light quarks. These hadrons may also be regarded to possess
quark-diquark structure \cite{kroll}.

Some authors \cite{lamdab1,Huang:1998rq,lamdab2,lamdab3} calculated
the form factors of $\Lambda_b$ decaying into light baryons and the
corresponding decay rates. The Ref. \cite{lamdab2} explored
$\Lambda_b \to p l \bar{\nu}$ by using the method of PQCD and they
concluded the perturbative analysis is reliable only for
$\rho(\equiv\frac{2p\cdot p'}{M^2_{\Lambda_b}})>0.8$. In Ref.
\cite{lamdab3}  the branching ratio of $\Lambda_b \to J/\Psi
\Lambda$ in PQCD was evaluated ($(1.7\sim5.3)\times
10^{-4}$)\cite{lamdab1}, instead, Cheng used the nonrelativistic
quark model to obtain this branching ratio as $1.1\times 10^{-4}$
which is lower than the experimental value ($(4.7\pm2.8)\times
10^{-4}$). In a recent study, the authors of \cite{Wang:2009hr} used
the light-cone sum rules to calculate the $\Lambda_b \to p(\Lambda)$
transition form factors.

In this work, we study the form factors of $\Lambda_b \to p$ and
$\Lambda_b \to \Lambda $ in the light-front model with the diquark
picture, and then we calculate the rates of $\Lambda_b \to p \pi$,
$\Lambda_b \to J/\Psi \Lambda$, as well as several other
non-leptonic decays of $\Lambda_b$.

When $\Lambda_b $ decays into light baryons, the energy of the light
baryon in the $\Lambda_b$ rest frame is
$E=(M^2_{\Lambda_b}+m^2-q^2)/(2M_{\Lambda_b})$ which is much larger
than its mass $m$ and the hadronic scale $\Lambda_{QCD}$. One
important feature of this region is that the light hadrons move
nearly along the light cone. It is argued in \cite{LEET} that the
active quark created from b quark by weak interaction carries most
of the energy of the final light baryon. Under the large energy
limit( LEET \cite{leet1}) and heavy quark limit( HQET \cite{Isgur})
we can obtain the relations between$f_3,g_3$ and $f_2,g_2$, which
may help to achieve the orders  of $f_3,g_3$. We write up these
relations in section II, and then derive the form factors
($f_1,f_2,g_1$ and $g_2$) of $\Lambda_b\to p$ and
$\Lambda_b\to\Lambda$ in section III. We carry out the numerical
computations in section IV. Finally, section V is devoted to
discussions by which we will draw our conclusion.

\section{Formulation}

\subsection{The form factors in the large energy limit} \label{sec}

The form factors for the weak transition $\Lambda_b\rightarrow H$
where $H$ represents a light baryon (refers to $p,~ \Lambda$ in this
study), are defined in the standard way as
\begin{eqnarray}\label{s1}
\mathcal{M}_{\mu}&=& \la H(P',S',S_z') \mid \bar q\gamma_{\mu}
 (1-\gamma_{5})b \mid \Lambda_{b}(P,S,S_z) \ra  \nonumber \\
 &=& \bar{u}_{H}(P',S'_z) \left[ \gamma_{\mu} f_{1}(q^{2})
 +i\sigma_{\mu \nu} \frac{ q^{\nu}}{M_{\Lambda_{b}}}f_{2}(q^{2})
 +\frac{q_{\mu}}{M_{\Lambda_{b}}} f_{3}(q^{2})
 \right] u_{\Lambda_{b}}(P,S_z) \nonumber \\
 &&-\bar u_{H}(P',S'_z)\left[\gamma_{\mu} g_{1}(q^{2})
  +i\sigma_{\mu \nu} \frac{ q^{\nu}}{M_{\Lambda_{b}}}g_{2}(q^{2})+
  \frac{q_{\mu}}{M_{\Lambda_{b}}}g_{3}(q^{2})
 \right]\gamma_{5} u_{\Lambda_{b}}(P,S_z),
\end{eqnarray}
where  $q \equiv P-P'$, $Q$ and $Q'$ denote heavy quark and light
quark,  $H$ stands as the light baryon, respectively. The above
formulation is the most general expression with only constraints of
enforcing the Lorentz invariance and parity conservation for strong
interaction. There are six form factors $f_i,~g_i$ (i=1,2,3) in
total for the vector and axial vector current $\bar
q\gamma_{\mu}(1-\gamma_5)b$ where the light-quark $q$ denotes $u$
for $p$ and $s$ for $\Lambda$. All the information about the strong
interaction is involved in those form factors. Since $S=S'=1/2$, we
will be able to write $\mid\Lambda_{b}(P,S,S_z)\ra$ as
$\mid\Lambda_{b}(P,S_z)\ra$ and similarly for $\bar u_{H}(P',S'_z)$
in the following formulations.

Another parametrization in terms of the four-velocities is widely
used and is found to be convenient for the heavy-to-heavy
transitions, such as $\Lambda_b\to\Lambda_c$. But for the
heavy-to-light transitions at the large recoil region where the
energy of final light baryon $H$ is much larger than its mass, it is
more convenient to use following formulation. Analogous to heavy
quark symmetry in heavy-to-heavy case, there is a large energy
symmetry relations for the heavy-to-light at large energy recoil
\cite{LEET}. For the heavy-to-light baryon transition, the symmetry
has not been searched up until present. In this subsection, we
explore the large-energy symmetry and show that they lead to a
simplification of the form factors: the six form factors are reduced
to three independent ones.

Let us introduce the velocity $v$ of initial $\Lambda_b$ and a light
front unit vector $n$ by
 \beq
 v=\frac{P}{M_{\Lambda_b}}, \qquad \qquad n=\frac{P'}{E},
 \eeq
where $E$ is the energy of $H$. Using these vectors, the amplitude
of the weak transition $\Lambda_b\rightarrow H$ is parameterized by
\begin{eqnarray} \label{ss2}
 \mathcal{M}_{\mu}&=&\la H(n,S_z') \mid \bar{q}\gamma_{\mu}
  (1-\gamma_{5})b \mid \Lambda_{Q}(v,S_z) \ra  \non \\
  &=& \bar{u}_{H}(n,S'_z)\left[F_1(E)\gamma_{\mu}
  +F_2(E)v_{\mu}+F_3(E)n_{\mu}\right]
  u_{\Lambda_{b}}(v,S_z)- \nonumber \\
 &&\bar u_{H}(n,S'_z)\left[G_1(E)\gamma_{\mu}
  +G_2(E)v_{\mu}+G_3(E)n_{\mu}\right]
 \gamma_{5} u_{\Lambda_{b}}(v,S_z).
 \end{eqnarray}
Up to leading order in $1/M_{\Lambda_b}$ the relation between the
two parametrization schemes  is
\begin{eqnarray}\label{gx}
&&f_1=F_1+\frac{1}{2}(\frac{F_2}{M_{\Lambda_b}}+\frac{F_3}{E})
M_{\Lambda_{b}},\qquad  g_1=G_1-\frac{1}{2}(\frac{G_2}
{M_{\Lambda_{b}}}+\frac{G_3}{E})M_{\Lambda_{b}},\nonumber\\
&&f_2=\frac{1}{2}(\frac{F_2}{M_{\Lambda_{b}}}+\frac{F_3}{E})M_{\Lambda_b},
\qquad\qquad\;
g_2=\frac{1}{2}(\frac{G_2}{M_{\Lambda_b}}+\frac{G_3}{E})M_{\Lambda_b},\non\\
&&f_3=\frac{1}{2}(\frac{F_2}{M_{\Lambda_b}}-\frac{F_3}{E})M_{\Lambda_b},
\qquad \qquad \;
g_3=\frac{1}{2}(\frac{G_2}{M_{\Lambda_{b}}}-\frac{G_3}{E})M_{\Lambda_{b}}.
\end{eqnarray}
where $M_{\Lambda_b}$ is the mass of $\Lambda_b$. We have neglected
the mass of final light baryon compared to $M_{\Lambda_b}$.

Under the large energy limit, the light energetic quark $q$ is
described by the two-component spinor
$\xi=\frac{\nslash\nbarslash}{4}q$ where $\bar n=2-n$ is another
light front unit vector and the heavy quark is replaced by
$h_v=e^{im_bv\cdot x}\frac{(1+\vslash)}{2}b$. The weak current $\bar
q\Gamma b$ in the full QCD is matched onto the current $\bar\xi
\Gamma h_v$ in the effective theory at tree level. For an arbitrary
matrix $\Gamma$, $\bar\xi \Gamma h_v$ has only three independent
Dirac structures. One convenient choice is discussed in
\cite{SCETff1}: $\bar\xi h_v$, $\bar\xi \gamma_5 h_v$ and $\bar\xi
\gamma_{\bot}^{\mu}h_v$. Thus, we have
 \beq
 \bar q\gamma^\mu b&=&\bar\xi\gamma_{\bot}^{\mu} h_v+n^{\mu}\bar\xi
 h_v,\non\\
 \bar q\gamma^\mu\gamma_5 b&=&i\epsilon_{\bot}^{\mu\nu}\bar\xi\gamma_{\bot}^{\nu} h_v
  -n^{\mu}\bar\xi \gamma_5 h_v.
 \eeq
where $\epsilon_\bot^{\mu\nu}=\epsilon^{\mu\nu\alpha\beta}v_\alpha
n_\beta$.

The three independent form factors are defined by
 \beq
 \la H(P',S_z')\mid \bar\xi h_v\mid \Lambda_b(P,S_z) \ra&=&
  \bar u_H(n,S_z)u_{\Lambda_b}(v,S_z)\zeta_0(E), \non\\
 \la H(P',S_z')\mid \bar\xi \gamma_5 h_v\mid \Lambda_b(P,S_z) \ra&=&
  \bar u_H(n,S_z)\gamma_5 u_{\Lambda_b}(v,S_z)\zeta_5(E), \non\\
 \la H(P',S_z')\mid \bar\xi\gamma_{\bot\mu} h_v\mid \Lambda_b(P,S_z) \ra&=&
  \bar u_H(n,S_z)\gamma_{\bot\mu}u_{\Lambda_b}(v,S_z)\zeta_{\bot}(E).
 \eeq

Then, we find
 \beq
 &&F_1=G_1=\zeta_\bot(E);  \qquad\qquad\;\;\;\; F_2=G_2=0;  \non\\
 &&F_3=\zeta_0(E)-\zeta_\bot(E);  \qquad\qquad
 G_3=\zeta_\bot(E)-\zeta_5(E).
 \eeq

From the above equation, we obtain the relations among the form
factors:
 \beq
 f_1+f_2=g_1-g_2; \qquad \qquad
 f_2=-f_3; \qquad\qquad g_2=-g_3.
 \eeq
This is one major result in this work. The $f_3$ and $g_3$ are not
independent, but related to $f_2$ and $g_2$.

\subsection{Vertex function in the light-front approach}

In the diquark picture, the heavy baryon $\Lambda_{b}$ is composed
of one heavy quark $b$ and a light diquark [ud]. In order to form a
color singlet hadron, the diquark [ud] is in a color anti-triplet.
Because $\Lambda_{b}$ is at the ground state, the diqaurk is a $0^+$
scalar ($s=0$, $l=0$) and the orbital angular momentum between the
diquark and the heavy quark is also zero, i.e. $L=l=0$. However the
situation is complicated  for light baryon even thought it is in the
ground state. The diquark in light baryon may be a $0^+$ scalar or a
$1^-$ vector. Fortunately  the diquark is a spectator in the
concerned transition and its spin is not affected so that only the
scalar diquark can transit into the final baryon and one only needs
to consider the scalar diquark structure of the light baryon.

In the light-front approach, the heavy baryon $\Lambda_Q$ composed
of only scalar diquark with total momentum $P$ and spin $S=1/2$
can be written as
\begin{eqnarray}\label{eq:lfbaryon}
 |\Lambda_Q(P,S,S_z)\rangle&=&\int\{d^3p_1\}\{d^3p_2\} \,
  2(2\pi)^3\delta^3(\tilde{P}-\tilde{p_1}-\tilde{p_2}) \nonumber\\
 &&\times\sum_{\lambda_1}\Psi^{SS_z}(\tilde{p}_1,\tilde{p}_2,\lambda_1)
  C_{\alpha\beta\gamma}F^{bc}\left|\right.
  Q^{\alpha}(p_1,\lambda_1)[q_{b}^{\beta}q_{c}^{\gamma}](p_2)\ra,
\end{eqnarray}
and  the light baryon (total momentum $P$, spin $J=1/2$, composed of
$0^+$ scalar diquark and orbital angular momentum $L=0$) has the
similar form,
\begin{eqnarray}
|H(P,S,S_z)\rangle&&=\int{d^3p_1}{d^3p_2} \,
2(2\pi)^3\delta^3(\tilde{P}-\tilde{p_1}-\tilde{p_2})
\nonumber\\&&\times
\sum_{\lambda_1}\Psi^{SS_z}(\tilde{p}_1,\tilde{p}_2,\lambda_1)
C_{\alpha,\beta,\gamma}F_{L}^{a,b,c}
|q_a^{\alpha}(p_1,\lambda_1)[q_{b}^\beta
q_{c}^\gamma](p_2)\rangle,
\end{eqnarray}
where $Q$, $[q_bq_c]$ represent heavy quark and diquark respectively
and $\lambda$ denotes the helicity, where $\alpha, \beta, \gamma$
and $a,b, c$ are the color and flavor indices, $p_1$, $p_2$ are the
on-mass-shell light-front momenta defined by
\begin{equation}
 \tilde{p}=(p^+,p_{\perp}),\qquad p_\perp=(p^1,p^2),\qquad
 p^-=\frac{m^2+p_{\perp}^2}{p^+},
\end{equation}
and
\begin{eqnarray}
&&\{d^3p\}\equiv\frac{dp^+d^2 p_{\perp}}{2(2\pi)^3},\qquad
  \delta^3(\tilde{p})=\delta(p^+)\delta^2(p_{\perp}),
  \nonumber\\
&&\mid Q(p_1,\lambda_1)[q_1 q_2](p_2)\rangle=
 b^{\dagger}_{\lambda_1}(p_1)a^{\dagger}(p_2)| 0\ra,\non\\
&&[a(p'), a^{\dagger}(p)]=2(2\pi)^3\delta^3(\tilde{p}'-\tilde{p}),
  \nonumber\\
&&\{d_{\lambda'}(p'),d_{\lambda}^{\dagger}(p)\}=
  2(2\pi)^3\delta^3(\tilde{p}'-\tilde{p})\delta_{\lambda'\lambda}.
\end{eqnarray}
The coefficient $C_{\alpha\beta\gamma}$ is a normalized color
factor and $F^{bc}(F^{abc})$ is a normalized flavor coefficient,
 \beq
 && C_{\alpha\beta\gamma}F^{bc}C_{\alpha'\beta'\gamma'}F^{b'c'}
  \la Q^{\alpha'}(p'_1,\lambda'_1)[q_{b'}^{\beta'}q_{c'}^{\gamma'}](p'_2)|
  Q^{\alpha}(p_1,\lambda_1)[q_{b}^{\beta}q_{c}^{\gamma}](p_2)\ra
  \non\\
  &&=2^2(2\pi)^6\delta^3(\tilde{p}_1'-\tilde{p}_1)\delta^3
  (\tilde{p}_2'-\tilde{p}_2)\delta_{\lambda'_1\lambda_1},\non\\
   && C_{\alpha\beta\gamma}F^{abc}C_{\alpha'\beta'\gamma'}F^{a'b'c'}
  \la q_{a'}^{\alpha'}(p'_1,\lambda'_1)[q_{b'}^{\beta'}q_{c'}^{\gamma'}](p'_2)|
  q_{a}^{\alpha}(p_1,\lambda_1)[q_{b}^{\beta}q_{c}^{\gamma}](p_2)\ra
  \non\\
  &&=2^2(2\pi)^6\delta^3(\tilde{p}_1'-\tilde{p}_1)\delta^3
  (\tilde{p}_2'-\tilde{p}_2)\delta_{\lambda'_1\lambda_1}.
 \eeq

In order to describe the motion of the constituents, one needs to
introduce intrinsic variables $(x_i, k_{i\perp})$ with $i=1,2$
through
\begin{eqnarray}
&&p^+_1=x_1 P^+, \qquad\qquad p^+_2=x_2 P^+,
 \qquad\qquad x_1+x_2=1, \nonumber\\
&&p_{1\perp}=x_1 P_{\perp}+k_{1\perp},
  ~~~ p_{2\perp}=x_2 P_{\perp}+k_{2\perp},
  ~~~ k_{\perp}=-k_{1\perp}=k_{2\perp},
\end{eqnarray}
where $x_i$'s are the light-front momentum fractions satisfying
$0<x_1, x_2<1$. The variables $(x_i, k_{i\perp})$ are independent of
the total momentum of the hadron and thus are Lorentz-invariant. The
invariant mass square $M_0^2$ is defined as
 \begin{eqnarray} \label{eq:Mpz}
  M_0^2=\frac{k_{1\perp}^2+m_1^2}{x_1}+
        \frac{k_{2\perp}^2+m_2^2}{x_2}.
 \end{eqnarray}
The invariant mass $M_0$ is in general different from the hadron
mass $M$ which satisfies the physical mass-shell condition
$M^2=P^2$. This is due to the fact that in the baryon, heavy quark
and diquark cannot be on their mass shells simultaneously. We define
the internal momenta as
 \beq
 k_i=(k_i^-,k_i^+,k_{i\bot})=(e_i-k_{iz},e_i+k_{iz},k_{i\bot})=
  (\frac{m_i^2+k_{i\bot}^2}{x_iM_0},x_iM_0,k_{i\bot}).
 \eeq
It is easy to obtain
 \begin{eqnarray}
  M_0&=&e_1+e_2, \non\\
  e_i&=&\frac{x_iM_0}{2}+\frac{m_i^2+k_{i\perp}^2}{2x_iM_0}
      =\sqrt{m_i^2+k_{i\bot}^2+k_{iz}^2},\non\\
 k_{iz}&=&\frac{x_iM_0}{2}-\frac{m_i^2+k_{i\perp}^2}{2x_iM_0}.
 \end{eqnarray}
where $e_i$ denotes the energy of the i-th constituent. The
momenta $k_{i\bot}$ and $k_{iz}$ constitute a momentum vector
$\vec k_i=(k_{i\bot}, k_{iz})$ and correspond to the components in
the transverse and $z$ directions, respectively.

In the momentum space, the  function $\Psi^{SS_z}$ appearing in Eq.
(\ref{eq:lfbaryon}) is expressed as
\begin{equation}
 \Psi^{SS_z}(\tilde{p}_1,\tilde{p}_2,\lambda_1)=
  \left\la\lambda_1\left|\mathcal{R}^{\dagger}_M(x_1,k_{1\perp},m_1)
   \right|s_1\right\ra
  \left\la 00;\frac{1}{2} s_1\left|\frac{1}{2}S_z\right\ra
   \phi(x,k_{\perp})\right.,
\end{equation}
where $\phi(x,k_{\perp})$  is the light-front wave function which
describes the momentum distribution of the constituents in the bound
state with $x=x_2,~k_{\perp}=k_{2\perp}$; and $\left\la
00;\frac{1}{2} s_1\left|\frac{1}{2}S_z\right\ra\right.$ is the
corresponding Clebsch-Gordan coefficient with total spin of the
scalar diquark  $s=s_z=0$;
$\left\la\lambda_1\left|\mathcal{R}^{\dagger}_M(x_1,k_{1\perp},m_1)
\right|s_1\right\ra$ is the well-known Melosh transformation matrix
element which transforms the the conventional spin states in the
instant form into the light-front helicity eigenstates,
 \beq
 \left\la\lambda_1\left|\mathcal{R}^{\dagger}_M(x_1,k_{1\perp},m_1)
 \right|s_1\right\ra &=& \frac{\bar u(k_1,\lambda_1)u_D(k_1,s_1)}{2m_1}
 \non\\
  &=&\frac{(m_1+x_1M_0)\delta_{\lambda_1 s_1}+i\vec{\sigma}_{\lambda_1 s_1}
  \cdot\vec k_{1\perp}\times\vec n}
  {\sqrt{(m_1+x_1M_0)^2+k_{1\perp}^2}},
 \eeq
where $u_{(D)}$ denotes a Dirac spinor in the light-front
(instant) form and $\vec n=(0,0,1)$ is a unit vector in the $z$
direction. In practice, it is more convenient to use the covariant
form for the Melosh transform matrix \cite{meson1,CCH2}
\begin{eqnarray}
 \left\la\lambda_1\left|\mathcal{R}^{\dagger}_M(x_1,k_{1\perp},m_1)
   \right|s_1\right\ra \left\la 00;\frac{1}{2}s_1\left|
   \frac{1}{2}S_z\right\ra\right.=\frac{1}{\sqrt{2(p_1\cdot
   \bar P+m_1M_0)}}\bar{u}(p_1,\lambda_1)\Gamma u(\bar {P},S_z),
\end{eqnarray}
where
\begin{eqnarray}
\Gamma=1, \qquad \qquad \bar {P}=p_1+p_2.
\end{eqnarray}
for the scalar diquark. If the diquark is a vector which is usually
supposed to be the case for the $\Sigma_{c(b)}$ baryon, the Melosh
transform matrix should be modified (since it is irrelevant to our
present work,  we omit the corresponding expressions).

The  baryon state is normalized as
 \beq
 \la
 \Lambda(P',S',S'_z)|\Lambda(P,S,S_z)\ra=2(2\pi)^3P^+
  \delta^3(\tilde{P}'-\tilde{P})\delta_{S'S}\delta_{S'_zS_z},
   \eeq
   the same for $H(P,S,S_z)$.

Thus, the light-front wave function obeys the constraint
 \beq
 \int\frac{dxd^2k_{\perp}}{2(2\pi^3)}|\phi(x,k_{\perp})|^2=1.
 \eeq

In principle, the wave functions can be obtained by solving the
light-front bound state equations. However, it is too hard to
calculate them based on the first principle, so that instead, we
would like to adopt a phenomenological function, and obviously, a
Gaussian form is most preferable,
 \beq
 \phi(x,k_{\perp})=N\sqrt{\frac{\partial k_{2z}}{\partial x_2}}
  {\rm exp}\left( \frac{-\vec k^2}{2\beta^2}\right).
 \eeq
with
 \beq
 N=4\left(\frac{\pi}{\beta^2}\right)^{3/4},\qquad
 \frac{\partial k_{2z}}{\partial x_2}=\frac{e_1e_2}{x_1x_2M_0}.
 \eeq
where $\beta$ determines the confinement scale. The phenomenological
parameters in the light-front quark model are quark masses and the
hadron wave function parameter $\beta$ which should be prior
determined before numerical computations can be carried out and we
will do the job in the later subsections.

\subsection{$\Lambda_{Q}\rightarrow H$ weak transitions}

Equipped with the light-front quark model description of $\mid
\Lambda_{Q}(P,S_z) >$ and $\mid H(P,S_z) >$, we can calculate the
weak transition matrix elements
\begin{eqnarray}\label{s2}
&& < \Lambda_{Q}(P',S_z') \mid \bar{q}
\gamma_{\mu} (1-\gamma_{5}) Q \mid H(P,S_z) >  \nonumber \\
 &=& N_{IF} \int\{d^3p_2\}\frac{\phi'^*_{H}(x',k'_{\perp})
 \phi_{\Lambda_{Q}}(x,k_{\perp})}{2\sqrt{p^+_1p'^+_1(p_1\cdot \bar{P}
 +m_1M_0)(p'_1\cdot \bar{P'}+m'_1M'_0)}}\nonumber \\&&
\times\bar{u}(\bar{P'},S'_z)\bar{\Gamma}'(p_1\!\!\!\!\!\slash'+m'_1)
\gamma_{\mu}(1-\gamma_{5})
(p_1\!\!\!\!\!\slash+m_1)\Gamma_{Lm}u(\bar{P},S_z),
\end{eqnarray}
where $N_{IF}$ is a flavor-spin factor of I (initial particle)
decaying into F (final particle). Following \cite{kroll}, the
flavor-spin functions of $\Lambda_b$, proton and $\Lambda$  take the
forms in the diquark picture
 \begin{eqnarray}\label{eq:flavor}
 && \chi_S^{\Lambda_b}=bS_{[u,d]}, \nonumber\\
 && \chi_S^{p}=uS_{[u,d]},\,\,\,
  \chi_V^{p}=[uV_{[u,d]}-\sqrt{2}dV_{[u,u]}  ]/\sqrt{3}    \nonumber\\
 &&
 \chi_S^{\Lambda}=[uS_{[d,s]}-dS_{[u,s]}-2sS_{[u,d]}]/\sqrt{6},\,\,\,\,\,
 \chi_V^{\Lambda}=[uV_{[d,s]}-dV_{[u,s]}]/\sqrt{2}
 \end{eqnarray}
where S and V denote scalar and axial vector diquark. We can get
$N_{\Lambda_b p}=\frac{1}{\sqrt{2}},N_{\Lambda_b
\Lambda}=\frac{1}{\sqrt{3}}$, which are consistent with
\cite{lamdab1}, and
 \beq
 &&\bar{\Gamma}'=\gamma_0\Gamma\gamma_0=\Gamma=1, \non \\
 &&m_1=m_b, \qquad m'_1=m_q, \qquad m_2=m_{[ud]}.
 \eeq
with $P$ and $P'$ denoting the momenta of initial and final baryons,
$p_1,~p'_1$ are the momenta of $b$ and $c$ quarks, respectively.
Because the diquark is a scalar, one does not need to deal with the
spinors which make computations more complex. In this framework, at
each effective vertex, only the three-momentum rather than the
four-momentum is conserved, hence
$\tilde{p}_1-\tilde{p}'_1=\tilde{q}$ and $\tilde{p}_2=\tilde{p}'_2$.
From $\tilde{p}_2=\tilde{p}'_2$, we have
 \beq
 x'=\frac{P^+}{P^{'+}}x, \qquad \qquad
 k'_{\perp}=k_{\perp}+x_2q_{\perp}.
 \eeq
with $x=x_2$, $x'=x'_2$. Thus, Eq. (\ref{s2}) is rewritten as
\begin{eqnarray}\label{s23}
 &&\la H(P',S_z') \mid \bar{q} \gamma^{\mu}
  (1-\gamma_{5}) Q \mid \Lambda_{Q}(P,S_z) \ra \non\\
  &=& N_{IF}\int\frac{dxd^2k_{\perp}}{2(2\pi)^3}\frac{
  \phi_{H}(x',k'_{\perp})
  \phi_{\Lambda_Q}(x,k_{\perp})}
  {2\sqrt{x_1x'_1(p_1\cdot \bar{P}+m_1M_0)
  (p'_1\cdot \bar{P'}+m'_1M'_0)}}\nonumber \\
 &&\times \bar{u}(\bar{P'},S'_z)
  (p_1\!\!\!\!\!\slash'+m'_1)\gamma^{\mu}(1-\gamma_{5})
  (p_1\!\!\!\!\!\slash+m_1) u(\bar{P},S_z).
\end{eqnarray}

Following \cite{befor,pentaquark1}, we get the the final expressions
for the $\Lambda_Q\to H$ weak transition form factors
\begin{eqnarray}\label{s9}
 f_1(q^2)&=&N_{IF}\int{\frac{dxd^2k_{\perp}}{2(2\pi)^3}}
   \frac{\phi_{H}(x',k'_{\perp})
  \phi_{\Lambda_Q}(x,k_{\perp})\left[k_{2\perp}
   \cdot k'_{2\perp}+\left(x_1M_0+m_1\right)
   \left(x'_1M'_0+m'_1\right)\right]}
   {\sqrt{\left[\left(m_1+x_1M_0\right)^2+k_{2\perp}^2\right]
   \left[\left(m'_1+x_1M'_0\right)^2+k_{2\perp}^{'2}\right]}}, \non \\
 g_1(q^2)&=&N_{IF}\int{\frac{dxd^2k_{\perp}}{2(2\pi)^3}}
   \frac{\phi_{H}(x',k'_{\perp})
  \phi_{\Lambda_Q}(x,k_{\perp})[-k_{2\perp}
   \cdot k'_{2\perp}+(x_1M_0+m_1)(x'_1M'_0+m'_1)]}
   {\sqrt{\left[\left(m_1+x_1M_0\right)^2+k_{2\perp}^2\right]
   \left[\left(m'_1+x_1M'_0\right)^2+k_{2\perp}^{'2}\right]}}, \non \\
 \frac{f_2(q^2)}{M_{\Lambda_Q}}&=&\frac{N_{IF}}{q^i_{\perp}}
   \int{\frac{dxd^2k_{\perp}}{2(2\pi)^3}}
   \frac{\phi_{H}(x',k'_{\perp})
   \phi_{\Lambda_Q}(x,k_{\perp})
   [(m_1+x_1M_0)k_{1\perp}^{\prime i}-(m'_1+x'_1M'_0)k_{1\perp}^i]}
   {\sqrt{\left[\left(m_1+x_1M_0\right)^2+k_{2\perp}^2\right]
   \left[\left(m'_1+x_1M'_0\right)^2+k_{2\perp}^{'2}\right]}}, \non \\
 \frac{g_2(q^2)}{M_{\Lambda_Q}}&=&\frac{N_{IF}}{q^i_{\perp}}
   \int{\frac{dxd^2k_{\perp}}{2(2\pi)^3}}
   \frac{\phi_{H}(x',k'_{\perp})
   \phi_{\Lambda_Q}(x,k_{\perp})
   [(m_1+x_1M_0)k_{1\perp}^{\prime i}+ (m'_1+x'_1M'_0)
   {k}_{1\perp}^i]}{\sqrt{\left[\left(m_1+x_1M_0\right)^2+
   k_{2\perp}^2\right]\left[\left(m'_1+x_1M'_0\right)^2+
   k_{2\perp}^{'2}\right]}}.\non\\
\end{eqnarray}
It is noted that the form factors $f_3$ and $g_3$ cannot be
extracted in our method because we have imposed the condition
$q^+=0$. The fact that the calculated $f_2$ and $g_2$ at $q^2=0$ are
small compared to $f_1$ and $g_1$ and the large energy limit
relations $f_3=-f_2$ and $g_3=-g_2$ show that using the large energy
limit relations for $f_3$ and $g_3$ does not produce substantial
theoretical errors.

\section{Semi-leptonic and Non-leptonic decays of transition
$\Lambda_b \to$ light hadrons }

In this section, we obtain formulations for the rates of
semi-leptonic and non-leptonic processes. In this work, we concern
only the exclusive decay modes.

\subsection{Semi-leptonic decays of $\Lambda_b \to p l\bar\nu_l$ }

Generally the polarization effects may be important for testifying
different theoretical models, so that we would pay more attention to
the physical consequences brought up by them. The transition
amplitude of $\Lambda_b\to p$ contains several independent helicity
components. According to the definitions of the form factors for
$\Lambda_b \to p$ given in Eq. (\ref{s1}), the helicity amplitudes
$H_{i,j}^V$ are related to these form factors through the following
expressions \cite{KKP}
 \beq
 H^V_{\frac{1}{2},0}&=&\frac{\sqrt{Q_-}}{\sqrt{q^2}}\left(
  \left(\Mb+\Mc\right)f_1-\frac{q^2}{\Mb}f_2\right),\non\\
 H^V_{\frac{1}{2},1}&=&\sqrt{2Q_-}\left(-f_1+
  \frac{\Mb+\Mc}{\Mb}f_2\right),\non\\
 H^A_{\frac{1}{2},0}&=&\frac{\sqrt{Q_+}}{\sqrt{q^2}}\left(
  \left(\Mb-\Mc\right)g_1+\frac{q^2}{\Mb}g_2\right),\non\\
 H^A_{\frac{1}{2},1}&=&\sqrt{2Q_+}\left(-g_1-
  \frac{\Mb-\Mc}{\Mb}g_2\right),
 \eeq
where $Q_{\pm}=2(P\cdot P'\pm \Mb M_p)=2\Mb M_p(\omega\pm 1)$.

The helicities of the $W$-boson $\lambda_W$ can be either $0$ or
$1$, corresponding to the longitudinal and transverse polarizations.
Following the definitions in literature, we decompose the decay
width into a sum of the longitudinal and transverse parts according
to the helicity states of the virtual W-boson. The differential
decay rate of $\Lambda_b \to p l\bar\nu_l$ is
 \beq
 \frac{d\Gamma}{d\omega}=\frac{d\Gamma_L}{d\omega}+
 \frac{d\Gamma_T}{d\omega},
 \eeq
and the longitudinally (L) and transversely (T) polarized rates are
respectively \cite{KKP}
 \beq
 \frac{d\Gamma_L}{d\omega}&=&\frac{G_F^2|V_{ub}|^2}{(2\pi)^3}~
  \frac{q^2~p_c~M_p}{12\Mb}\left[
  |H_{\frac{1}{2},0}|^2+|H_{-\frac{1}{2},0}|^2\right],\non\\
 \frac{d\Gamma_T}{d\omega}&=&\frac{G_F^2|V_{ub}|^2}{(2\pi)^3}~
  \frac{q^2~p_c~M_p
}{12\Mb}\left[
  |H_{\frac{1}{2},1}|^2+|H_{-\frac{1}{2},-1}|^2\right].
 \eeq
where $p_c=M_p\sqrt{\omega^2-1}$ is the momentum of the proton in
the rest frame of $\Lambda_b$. The relations between $H_{i,j}$ and
$H^V_{i,j}$ can be found in \cite{KKP}. Integrating over the solid
angle, we obtain the decay rate as
 \beq
 \Gamma=\int_1^{\omega_{\rm max}}d\omega\frac{d\Gamma}{d\omega},
 \eeq
where the upper bound of the integration $\omega_{\rm
max}=\frac{1}{2}\left(\frac{M_{\Lambda_{b}}}
{M_{p}}+\frac{M_{p}}{M_{\Lambda_{b}}}\right)$ corresponds to the
maximal recoil. In order to compare our results with those in the
literatures, we use the variable $\omega$ in the expression for the
differential decay rate.

The polarization of the cascade decay $\Lambda_b\to p+W(\to l\nu)$
is expressed by various asymmetry parameters \cite{EFG,KKP}. Among
them, the integrated longitudinal and transverse asymmetries are
defined by
 \beq
 a_L&=&\frac{\int_1^{\omega_{\rm max}} d\omega ~q^2~ p_c
     \left[ |H_{\frac{1}{2},0}|^2-|H_{-\frac{1}{2},0}|^2\right]}
     {\int_1^{\omega_{\rm max}} d\omega ~q^2~ p_c
     \left[|H_{\frac{1}{2},0}|^2+|H_{-\frac{1}{2},0}|^2\right]},
     \non\\
 a_T&=&\frac{\int_1^{\omega_{\rm max}} d\omega ~q^2~ p_c
     \left[ |H_{\frac{1}{2},1}|^2-|H_{-\frac{1}{2},-1}|^2\right]}
     {\int_1^{\omega_{\rm max}} d\omega ~q^2~ p_c
     \left[|H_{\frac{1}{2},1}|^2+|H_{-\frac{1}{2},-1}|^2\right]}.
 \eeq
The ratio of the longitudinal to transverse decay rates $R$ is
defined by
 \beq
 R=\frac{\Gamma_L}{\Gamma_T}=\frac{\int_1^{\omega_{\rm
     max}}d\omega~q^2~p_c\left[ |H_{\frac{1}{2},0}|^2+|H_{-\frac{1}{2},0}|^2
     \right]}{\int_1^{\omega_{\rm max}}d\omega~q^2~p_c
     \left[ |H_{\frac{1}{2},1}|^2+|H_{-\frac{1}{2},-1}|^2\right]},
 \eeq
and the  longitudinal proton polarization asymmetry $P_L$ is given
as
 \beq
 P_L&=&\frac{\int_1^{\omega_{\rm max}} d\omega ~q^2~ p_c
     \left[ |H_{\frac{1}{2},0}|^2-|H_{-\frac{1}{2},0}|^2+
     |H_{\frac{1}{2},1}|^2-|H_{-\frac{1}{2},-1}|^2\right]}
     {\int_1^{\omega_{\rm max}} d\omega ~q^2~ p_c
     \left[|H_{\frac{1}{2},0}|^2+|H_{-\frac{1}{2},0}|^2+
     |H_{\frac{1}{2},1}|^2+|H_{-\frac{1}{2},-1}|^2\right]}
  \non\\
  &=&\frac{a_T+R\,a_L}{1+R}.
 \eeq

\subsection{Non-leptonic decay  of $\Lambda_b \to p \,+M$}

From the theoretical aspects, the non-leptonic decays are much
more complicated than the semi-leptonic ones because of the
strong interaction. Generally, the present theoretical framework
is based on the factorization assumption, where  the hadronic
matrix element is factorized into a product of two matrix
elements of single currents. One can be written as a decay
constant while the other is expressed in terms of a few form
factors according to the lorentz structure of the current. For the
weak decays of mesons, such factorization approach is verified to
work very well for the color-allowed processes and the
non-factorizable contributions are negligible.

For the non-leptonic decays $\Lambda_b^0 \to p+ M$, the effective
interaction at the quark level is $b\to u\bar{q_1}q_2$. The
relevant Hamiltonian is
 \beq
 &&{\cal H}_W=\frac{G_F}{\sqrt 2}V_{ub}V_{q_1q_2}^*(c_1O_1+c_2O_2),
  \non\\
 &&O_1=(\bar u b)_{V-A} (\bar q_2q_1)_{V-A},\qquad
 O_2=(\bar q_2b)_{V-A} (\bar u q_1)_{V-A}.
 \eeq
where $c_i$ denotes the short-distance Wilson coefficient,
$V_{ub}(V_{q_1q_2})$ is the CKM matrix elements, $q_1$ stands for
$u$  and $q_2$ for $d$  in the context. Then one needs to evaluate
the hadronic matrix elements
 \beq
 \la p\,M | {\cal H}_W | \Lambda_b\ra=
 \frac{G_F}{\sqrt 2}V_{ub}V_{q_1q_2}^*\sum_{i=1,2}c_i~
 \la p\,M | O_i | \Lambda_b\ra.
 \eeq
Under the factorization approximation, the hadronic matrix element
is reduced to
 \beq
 \la p\,M | O_i | \Lambda_b\ra
 =\la p | J_\mu |\Lambda_b\ra
  \la M | J^{\prime\mu} | 0\ra.
 \eeq
where $J(J')$ is the $V-A$ weak current. The first factor $\la p |
J_\mu |\Lambda_b\ra$ is parameterized by six form factors as done
in Eq. (\ref{s1}). The second factor defines the decay constants
as follows
 \beq\label{p1}
 \la P(P)|A_{\mu}|0\ra&=&f_PP_{\mu}, \non\\
 \la S(P)|V_{\mu}|0\ra&=&f_SP_{\mu}, \non\\
 \la V(P,\epsilon)|V_{\mu}|0\ra&=&f_VM_V\epsilon^*_{\mu}, \non\\
 \la A(P,\epsilon)|A_{\mu}|0\ra&=&f_VM_A\epsilon^*_{\mu},
 \eeq
where $P(V)$ denotes a pseudoscalar (vector) meson, and $S(A)$
denotes a scalar (axial-vector) meson. In the definitions, we omit
a factor $(-i)$ for the  pseudoscalar meson decay constant.

In general, the transition amplitude of $\Lambda_b\to p \pi^-$ can
be written as
 \beq\label{p2}
 {\cal M}(\Lambda_b\to p P)&=&\bar
  u_{p}(A+B\gamma_5)u_{\Lambda_b}, \non \\
 {\cal M}(\Lambda_b\to p V)&=&\bar
  u_{p}\epsilon^{*\mu}\left[A_1\gamma_{\mu}\gamma_5+
   A_2(p_{\Lambda_c})_{\mu}\gamma_5+B_1\gamma_{\mu}+
   B_2(p_{\Lambda_c})_{\mu}\right]u_{\Lambda_b},
 \eeq
where $\epsilon^{\mu}$ is the polarization vector of the final
vector or axial-vector mesons. Including the effective Wilson
coefficient $a_1=c_1+c_2/N_c$, the decay amplitudes under the
factorization approximation are \cite{KK,Cheng} \beq\label{p3}
 A&=&\lambda f_P(\Mb-\Mc)f_1(M^2), \non \\
 B&=&\lambda f_P(\Mb+\Mc)g_1(M^2), \non\\
 A_1&=&-\lambda f_VM\left[g_1(M^2)+g_2(M^2)\frac{\Mb-\Mc}{\Mb}\right],
 \non\\
 A_2&=&-2\lambda f_VM\frac{g_2(M^2)}{\Mb},\non\\
 B_1&=&\lambda f_VM\left[f_1(M^2)-f_2(M^2)\frac{\Mb+\Mc}{\Mb}\right],
 \non\\
 B_2&=&2\lambda f_VM\frac{f_2(M^2)}{\Mb},
 \eeq
where $\lambda=\frac{G_F}{\sqrt 2}V_{ub}V_{q_1q_2}^*a_1$ and $M$ is
the $\pi$ mass. Replacing  $P$, $V$ by $S$ and $A$ in the above
expressions, one can easily obtain similar expressions for scalar
and axial-vector mesons.

The decay rates of $\Lambda_b\rightarrow p\pi^-$ and up-down
asymmetries are \cite{Cheng}
 \begin{eqnarray}\label{p4}
 \Gamma&=&\frac{p_c}{8\pi}\left[\frac{(\Mb+M_p)^2-M^2}{\Mb^2}|A|^2+
  \frac{(\Mb-M_p)^2-M^2}{\Mb^2}|B|^2\right], \non\\
 \alpha&=&-\frac{2\kappa{\rm Re}(A^*B)}{|A|^2+\kappa^2|B|^2},
 \end{eqnarray}
where $p_c$ is the proton momentum in the rest frame of
$\Lambda_b$ and $\kappa=\frac{p_c}{E_{p}+M_p}$. For
$\Lambda_b\rightarrow\Lambda_c V(A)$ decays, the decay rates and
up-down asymmetries are
 \beq\label{p5}
 \Gamma&=&\frac{p_c (E_{p}+M_p)}{8\pi\Mb}\left[
  2\left(|S|^2+|P_2|^2\right)+\frac{E^2}{M^2}\left(
  |S+D|^2+|P_1|^2\right)\right], \non\\
 \alpha&=&\frac{4M^2{\rm Re}(S^*P_2)+2E^2{\rm Re}(S+D)^*P_1}
  {2M^2\left(|S|^2+|P_2|^2 \right)+E^2\left(|S+D|^2+|P_1|^2
  \right) },
 \eeq
where $E$ is the energy of the vector (axial vector) meson, and
 \begin{eqnarray}\label{p6}
  S&=&-A_1, \non\\
  P_1&=&-\frac{p_c}{E}\left(\frac{\Mb+M_p}
  {E_{p}+M_p}B_1+B_2\right), \non \\
  P_2&=&\frac{p_c}{E_{p}+M_p}B_1,\non\\
  D&=&-\frac{p^2_c}{E(E_{p}+M_p)}(A_1- A_2).
 \end{eqnarray}

\subsection{Non-leptonic decay $\Lambda_b \rightarrow \Lambda+M $}

Theses decays proceed only via the internal W-emission. With the
factorization assumption, the amplitude is
 \beq
 A( \Lambda_b\rightarrow  \Lambda\,M)=
 \frac{G_F}{\sqrt 2}V_{qb}V_{q's}^*a_2~
 \la M| \bar{q'}\gamma_\mu(1-\gamma_5)q |0\ra\la\Lambda|\bar{s}
 \gamma^\mu(1-\gamma_5)b| \Lambda_b\ra.
 \eeq

In general, we can use the same formula
(Eqs.(\ref{p1})-(\ref{p6})) to obtain the decay rates and up-down
asymmetries of $\Lambda_b\to\Lambda+ M$. Note that: (1) at this
time $\lambda$ is replaced by $\frac{G_F}{\sqrt
2}V_{ub}V_{q_1q_2}^*a_2$, (2) when q and $\bar{q'}$ are u and
$\bar{u}$ respectively, the final meson may be $\pi^0$, $\eta$ or
$\eta'$.

For the decay constants of $\pi^0$, $\eta$ and $\eta'$, we have
 \beq
 \la \pi^0|\bar{u}\gamma_\mu\gamma_5u|0\ra&=&f^u_{\pi^0}P_{\mu}, \non\\
 \la \eta|\bar{u}\gamma_\mu\gamma_5u|0\ra&=&f^u_{\eta}P_{\mu}, \non\\
 \la \eta'|\bar{u}\gamma_\mu\gamma_5u|0\ra&=&f^u_{\eta'}P_{\mu},
 \eeq
where$f^u_{\pi^0}=\frac{f_{\pi}}{\sqrt{2}}$, $f^u_{\eta}$ and
$f^u_{\eta'}$ can be get form \cite{mix}.

\section{Numerical Results}

In this section we perform the numerical computations of the form
factors for $\Lambda_b \rightarrow p$ and $\Lambda_b \rightarrow
\Lambda$, then using them we estimate the  rates of
$\Lambda_b\rightarrow p+l\,\nu$, $\Lambda_b\rightarrow p+ M$ and
$\Lambda_b\rightarrow \Lambda+M$ where $M$ stands as various mesons.

In our calculation, the quark masses of $m_b$ and $m_s$ are taken
from \cite{pentaquark1}; $m_u$ is set to be 0.3 GeV; the mass of
diquark [ud], parameters $\beta_{b,[ud]}$, $\beta_{s,[ud]}$ and
$\beta_{u,[ud]}$ are chosen from \cite{befor,CCH2,pentaquark1}. The
baryon masses $M_{\Lambda_b}=5.624$ GeV, $M_p=0.938$ GeV, $
\Lambda=1.116$ GeV come from \cite{PDG06}. The input parameters are
collected in Table \ref{t3}.

\begin{table}\label{t3}
\caption{Input parameters in LFQM (in units of GeV).}
\begin{center}
\begin{ruledtabular}
\begin{tabular}{ccccccc}
 $m_b$ & $m_s$ & $m_u$ & $m_{[ud]}$ & $\beta_{u,[ud]}$ & $\beta_{b,[ud]}$
  & $\beta_{s,[ud]}$ \\\hline
 4.4  & 0.45  & 0.3  & 0.5 & 0.3 & 0.4 & 0.3
\end{tabular}
\end{ruledtabular}
\end{center}
\end{table}

\subsection{Form factor}

\begin{figure}
\begin{center}
\scalebox{0.7}{\includegraphics{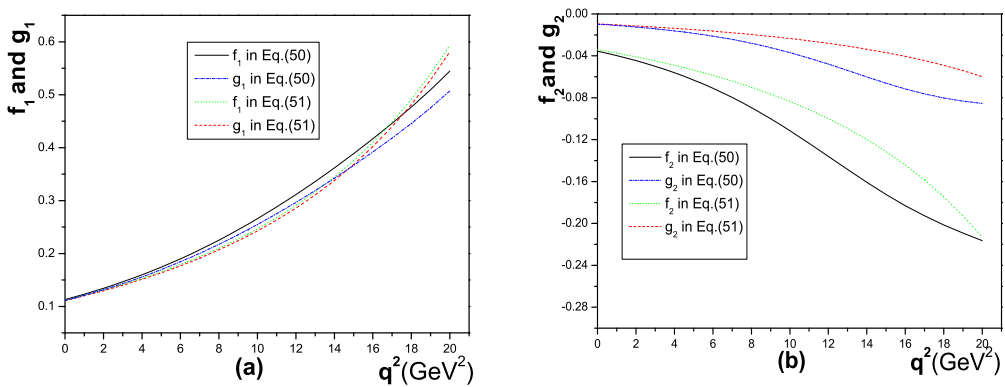}}
\end{center}
\caption{(a) Form factors  $ f_1$ and $g_1$ of $\Lambda_b\to p$.
\,\,\,   (b) Form factors  $ f_2$ and $g_2$ of $\Lambda_b\to p$.}
\label{t1}
\end{figure}

In LFQM, the calculation of form factors is performed in the frame
$q^+=0$ with $q^2=-q^2_{\perp}\leq 0$, only the values of the form
factors in the space-like region can be obtained. The advantage of
this choice is that the so-called Z-graph contribution arising from
the non-valence quarks vanishes. In order to obtain the physical
form factors, an extrapolation from the space-like region to the
time-like region is required. Following \cite{pentaquark1}, the form
factors in the space-like region can be parameterized in a
three-parameter form as
 \begin{eqnarray}\label{s14}
 F(q^2)=\frac{F(0)}{\left(1-\frac{q^2}{M_{\Lambda_b}^2}\right)
  \left[1-a\left(\frac{q^2}{M_{\Lambda_b}^2}\right)
  +b\left(\frac{q^2}{M_{\Lambda_b}^2}\right)^2\right]},
 \end{eqnarray}
where $F$ represents the form factor $f_{1,2}$ and $g_{1,2}$. The
parameters $a,~b$ and $F(0)$ are fixed by performing a
three-parameter fit to the form factors in the space-like region
which were obtained in previous sections. We then use these
parameters to determine the physical form factors in the time-like
region. The fitted values of $a,~b$ and $F(0)$ for different form
factors $f_{1,2}$ and $g_{1,2}$ are given in Table \ref{Tab:t2}
and \ref{Tab:t3}. The $q^2$ dependence of the form factors is
plotted in Fig. \ref{t1}.

\begin{table}
\caption{The $\Lambda_b\to p$ form factors  in the
  three-parameter form.}\label{Tab:t2}
\begin{ruledtabular}
\begin{tabular}{cccc}
  $F$    &  $F(0)$   &  $a$   &  $b$   \\
  $f_1$  &  0.1131   &  1.70  &  1.60  \\
  $f_2$  &  -0.0356  &  2.50  &  2.57  \\
  $g_1$  &  0.1112   &  1.65  &  1.60  \\
  $g_2$  &  -0.0097  &  2.80  &  2.70
\end{tabular}
\end{ruledtabular}
\end{table}

\begin{table}
\caption{The $\Lambda_b\to \Lambda$ form factors in the
  three-parameter form.}\label{Tab:t3}
\begin{ruledtabular}
\begin{tabular}{cccc}
  $F$    &  $F(0)$ &  $a$  &  $b$ \\
  $f_1$  &   0.1081    &   1.70    & 1.60  \\
  $f_2$  &   -0.0311     &   2.50    &  2.50  \\
  $g_1$  &    0.1065   &     1.70  &  1.40  \\
  $g_2$  &      -0.0064   &     2.70  & 2.70
\end{tabular}
\end{ruledtabular}
\end{table}

From Fig. \ref{t1}, we can see that there is only a tiny difference
between $f_1$ and $g_1$, i.e. they are close to each other. $g_2$ is
small comparing to $f_1$ and $g_1$. This is the same as the
conclusion of \cite{befor,peng}. But the difference between $f_2$
and $g_2$ increases as $q^2$ increases. This will break the large
energy limit relation $f_1+f_2=g_1-g_2$ proposed in the Section
\ref{sec}.

Our method of smooth extrapolation of from factors from space- to
time-like momentum regions is by no means an analytical continuation
in the rigorously mathematical sense but an extension, although it
is used in many phenomenological analysis. In
\cite{Becirevic:1999kt}, the authors suggest to write the form
factor as a dispersion relation in $q^2$ with a lowest-lying pole
plus a contribution from multiparticle states. We follow this scheme
and use a parametrization method adopted in \cite{Ball:2004ye}
 \begin{eqnarray}\label{s14p}
 F(q^2)=\frac{r_1}{\left(1-\frac{q^2}{M_{fit}^2}\right)}+
 \frac{r_2}{\left(1-(\frac{q^2}{M_{fit}^2})^2\right)},
  \end{eqnarray}
The parameters $r_1,~r_2$ and $M_{fit}$ are fixed in the space-like
regions for the transition of $\Lambda_b \rightarrow p$. The results
are presented in Table \ref{Tab:t2p}. We also plot the form factors
in the new parametrization method in Fig. \ref{t1} for a comparison.
From Fig. \ref{t1},  we can find there ia a little difference
between the form factors fitted by the above two methods. In
particular, the $f_1$ and $g_1$ in the two methods are nearly the
same. The difference of $f_2$ and $g_2$ in the methods increases
when $q^2$ increases, but due to smallness of their values, they
will not produce substantial errors to our predictions.

\begin{table}
\caption{The $\Lambda_b\to p$ form factors  in the
  the form of Eq.(\ref{s14p}).}\label{Tab:t2p}
\begin{ruledtabular}
\begin{tabular}{cccc}
  $F$    &  $r_1$   &  $r_2$   &  $M_{fit}$   \\
  $f_1$  &  -0.183   &  0.295  &  6.8  \\
  $f_2$  &  -0.176  &  0.287  &  6.8  \\
  $g_1$  &  0.080   &  -0.114  &  6.8  \\
  $g_2$  &  0.023  &  -0.032  &  6.8
\end{tabular}
\end{ruledtabular}
\end{table}

\subsection{Semi-leptonic decay of $\Lambda_b \to
p +l\bar{\nu}_l$}

With the form factors given in the above subsection, we are able to
calculate the branching ratio and various asymmetries of $\Lambda_b
\to p l\bar{\nu}_l$ decay. Table \ref{Tab:t4} presents our numerical
predictions. The ratio of longitudinal to transverse rates $R>1$
implies that the longitudinal polarization dominates.

\begin{table}
\caption{The branching ratios and polarization asymmetries of
$\Lambda_b\to p l\bar{\nu}_l$ .}\label{Tab:t4}
\begin{ruledtabular}
\begin{tabular}{ccccc}
    $BR$                 &  $a_L$  &  $a_T$ & $R$   &  $P_L$ \\\hline
   $2.54\times 10^{-4}$  & -0.99  & -0.96 & 1.11 & -0.97
\end{tabular}
\end{ruledtabular}
\end{table}

\subsection{Non-leptonic decays of $\Lambda_b\to p+M$ and $\Lambda_b\to \Lambda+M$}

The non-leptonic decays $\Lambda_b\to p(\Lambda)+M$ in the
factorization approach have been studied in the previous section.
Now, we present our numerical predictions on the decay rates and
relevant measurable quantities. The CKM matrix elements take the
values \cite{PDG06}
 \beq
 && V_{ud}=0.97377, \qquad V_{us}=0.2257, \qquad V_{cd}=0.230, \qquad
 \nonumber\\
 && V_{cs}=0.957,~~~\qquad V_{cb}=0.0416, \qquad ~V_{ub}=0.00413, \qquad
 \eeq
and the effective Wilson coefficient $a_1= 1$ \cite{pentaquark1},
$a_2=0.23$ \cite{lamdab1}. The meson decay constants are shown in
Table \ref{Tab:t5}.

\begin{table}
\caption{Meson decay constants $f$ (in units of
 MeV) \cite{CCH2, lamdab1}.}\label{Tab:t5}
\begin{ruledtabular}
\begin{tabular}{ccccccccccc}
  meson & $\pi$ & $\rho$ & $K$ & $K^*$ & $D$ & $D^*$ & $D_s$ & $D_s^*$ &
  $a_1$& $J/\psi$ \\\hline
  $f$   & 131   & 216    & 160 & 210   & 200 & 220   & 230   & 230     &
  203& 395
\end{tabular}
\end{ruledtabular}
\end{table}
The predictions for $\Lambda_b^0\to p+M$ are provided in Table
\ref{Tab:t7}. The Table \ref{Tab:t8} demonstrates a comparison of
our result with  other approaches and experimental data for
$\Lambda_b^0\to\Lambda J/\psi$. In the Table \ref{Tab:t9} we give
predictions on the rates of $\Lambda_b^0\to \Lambda+$meson.

\begin{table}
\caption{Branching ratio and up-down asymmetry for non-leptonic
decay $\Lambda_b^0\to p+M$.}\label{Tab:t7}
\begin{ruledtabular}
\begin{tabular}{cccccccc}
  & Branching ratios &Up-down asymmetries & Exp\\\hline
 $\Lambda_b^0\to p+ \pi^-$   & $3.15\times 10^{-6}$ &-1 &  $(3.5\pm 0.6(stat)\pm 0.9(syst))\times 10^{-6}$\\\hline
$\Lambda_b^0\to p+ \rho$   & $6.12\times 10^{-6}$ &-0.873 & $-$
\\\hline $\Lambda_b^0\to p+ a_1$   & $4.08\times 10^{-6}$
&-0.741 & $-$ \\\hline $\Lambda_b^0\to p+ D^-$   & $5.75\times
10^{-7}$ &-0.998 & $-$ \\\hline $\Lambda_b^0\to p+ D^{*-}$   &
$6.05\times 10^{-7}$ &-0.546\\\hline $\Lambda_b^0\to p+ D_s$   &
$1.36\times 10^{-5}$ &-0.997 & $-$\\\hline $\Lambda_b^0\to p+
D_s^*$ & $6.70\times 10^{-6}$ &-0.514 & $-$
\\\hline$\Lambda_b^0\to p+ K$ & $2.58\times 10^{-7}$ &-1 &
$(5.8\pm 0.8(stat)\pm 1.5(syst))\times 10^{-6}$
\\\hline $\Lambda_b^0\to p+ K^*$   & $3.21\times 10^{-7}$ &-0.850 &
$-$
\end{tabular}
\end{ruledtabular}
\end{table}

\begin{table}[!h]
\caption{Branching ratio and up-down asymmetry for non-leptonic
decay $\Lambda_b^0\to\Lambda  J/\psi$ within different theoretical
approaches and data from experiment .}\label{Tab:t8}
\begin{ruledtabular}
\begin{tabular}{cccccccc}
  & This work    &\cite{lamdab3} & \cite{Cheng} & \cite{MGKIIO}  & \cite{FR}
  & Exp.\cite{PDG06}       \\\hline
  Br($\times 10^{-4}$) & $3.94$ & $1.65\sim 5.27$    &$1.6$  & $2.55$  &
  $6.037$&$4.7\pm2.8$  \\\hline
  $\alpha$ & -0.204 &$-0.17\sim -0.14$&-0.1  &  -0.208& -0.18&-
\end{tabular}
\end{ruledtabular}
\end{table}

\begin{table}
\caption{Branching ratio and up-down asymmetry for non-leptonic
decay $\Lambda_b^0\to \Lambda+M$ with different theoretical
approaches .}\label{Tab:t9}
\begin{ruledtabular}
\begin{tabular}{ccccccc}
  & Branching ratios &Up-down asymmetries \\\hline
 $\Lambda_b^0\to \Lambda+\pi^0$    & $7.49\times 10^{-8}$ &-1  \\\hline
 $\Lambda_b^0\to \Lambda+\eta$     & $5.46\times 10^{-8}$ &-1 &  \\\hline
 $\Lambda_b^0\to \Lambda+\eta'$    & $2.29\times 10^{-8}$ &-1  \\\hline
 $\Lambda_b^0\to \Lambda+D^0$      & $4.54\times 10^{-5}$ &-0.998 \\\hline
 $\Lambda_b^0\to \Lambda+D^{0*}$   & $4.78\times 10^{-5}$ &-0.551 \\\hline
 $\Lambda_b^0\to \Lambda+\bar{D}^0$& $8.76\times 10^{-6}$ &-0.998 \\\hline
 $\Lambda_b^0\to \Lambda+\bar{D}^{0*}$ & $5.08\times 10^{-6}$ &-0.551
\end{tabular}
\end{ruledtabular}
\end{table}

From  Table \ref{Tab:t8} we can find that there are some differences
among the predictions by various theoretical approaches. In our
calculation, the $f_1(m^2_{J/\psi}),g_1(m^2_{J/\psi})$  is nearly
equal, however  $g_1(m^2_{J/\psi})$ is bigger than
$f_1(m^2_{J/\psi})$ in \cite{lamdab1,Cheng}.

\section{Conclusion}

In this work, we carefully investigate the processes where a heavy
baryon decays into a light baryon plus a lepton pair (semi-leptonic
decay) or a meson (non-leptonic decay) in terms of the
light-front-quark model(LFQM). Besides the regular input parameters
such as the quark masses and well measured decay constants of
various mesons, there is only one free parameter to be determined,
that is $\beta$ in the light front wavefunction. In our earlier work
\cite{befor}, by fitting the data of the semi-leptonic decays
$\Lambda_b\to\Lambda_c+l+\bar\nu$, we obtained the values of
$\beta_{b[ud]}$. Similarly, we fix the values $\beta_{u,[ud]}$ for
proton and $\beta_{s,[ud]}$ for $\Lambda$.

Our numerical results are shown in corresponding tables and some
measurable quantities such as the up-down asymmetries are also
evaluated. A clear comparison of our prediction on the decay rate of
$\Lambda_b\rightarrow \Lambda+J/\psi$ with the results predicted by
other models and as well as the experimental data is also explicitly
presented. One can notice that our result for
$\Lambda_b\to\Lambda+J/\psi$ is $3.94\times 10^{-4}$ which is in
good agreement with the data. The success is not too surprising even
though the model we adopt is much simplified. Definitely this value
obtained in this work is closer to the cental value of measurement
than the previous evaluations, but since there is a large
uncertainty in the data, one still cannot justify which model is
more preferable than others because within two standard deviations,
all the numerical results achieved with all the approaches listed in
the table are consistent with data. The asymmetry parameter which
may be important for determining the applicability of the adopted
model, is estimated as $-0.204$, which is generally consistent with
that obtained in other models and approaches. Of course the details,
especially the branching ratios will be further tested by the more
accurate experiments in the future.

Besides the semi-leptonic decays, we also estimate the branching
ratios of several non-leptonic decay modes which are listed in Table
VIII. Recently the CDF collaboration\cite{CDF2009} has measured the
branching ratios of $\Lambda_b\to p+\pi^-$ and $\Lambda_b\to p+K^-$
as $BR(\Lambda_b\to p+\pi^-)=(3.5\pm 0.6(stat)\pm 0.9(syst))\times
10^{-6}$ and $BR(\Lambda_b\to p+K^-)=(5.8\pm 0.8(stat)\pm
1.5(syst))\times 10^{-6}$. Our prediction on $BR(\Lambda_b\to
p+\pi^-)$=($3.15\times 10^{-6}$) is consistent with the measurement
of the CDF within one standard deviation, but for $BR(\Lambda_b\to
p+K^-)$ our value is $2.58\times 10^{-7}$, one order smaller than
the data of the CDF collaboration. Following the literature, in our
calculation, we employ the factorization scheme where the emitted
pseudoscalar meson ($\pi$ or K) is factorized out and described by
the common-accepted form factor $<0|A_{\mu}|M>=if_M p_{\mu}$ where
$A_{\mu},\ f_M$ and $p_{\mu}$ are the corresponding axial current,
decay constant of meson M and its four-momentum respectively. It is
noticed that in the case $\Lambda_b\to p+\pi^-$, at the vertex
$W^-\bar ud$ the Cabibbo-Kabayashi-Maskawa entry is approximately
$\cos\theta_C\approx 1$ whereas for the case $\Lambda_b\to p+K^-$,
the CKM entry is $\sin\theta_C\approx 0.22$, thus comparing with
$\Lambda_b\to p+\pi^-$, the amplitude of the process $\Lambda_b\to
p+K^-$ is suppressed by a factor ${f_K\over f_\pi}\sin\theta_C\sim
0.27$. Thus besides a small difference between the final phase
spaces of the two reactions, one can roughly estimate that
$BR(\Lambda_b\to p+K^-)/BR(\Lambda_b\to p+\pi^-)\sim
 0.07$, and this estimate is consistent with our
numerical results. Therefore the smallness of $BR(\Lambda_b\to
p+K^-)$ seems reasonable. However the data of CDF show completely
different results that $BR(\Lambda_b\to p+K^-)$ is anomalously
larger than $BR(\Lambda_b\to p+\pi^-)$.

In fact, in our calculations on the non-leptonic decays, we only
consider the contributions from the tree diagrams and neglect the
penguin-loop effects. For the mode of $\Lambda_b\to p+\pi^-$, the
penguin contribution can be safely neglected compared to the tree
level. However, for the mode of $\Lambda_b\to p+K^-$, the tree level
contribution is suppressed by the Cabibbo-Kabayashi-Maskawa entry
$V_{ub}V_{us}^*$ while for the penguin diagram, the main
contribution comes from the loop where top quark is the intermediate
fermion. In the case, the CKM entry would be $V_{tb}V_{ts}^*$ which
is almost two orders larger than $V_{ub}V_{us}^*$. Thus even though
there is a loop suppression of order $\alpha_s/4\pi$, it is
compensated by the much larger CKM entry. This situation was
discussed in \cite{Wangym} where the authors used the pQCD method to
carry out the calculations. In fact, we make a rough estimation of
the contribution from the top-penguin, and the result is almost five
times larger than the contribution from the tree diagram given
above.

However, from another aspect, when the penguin diagram is taken into
account, the factorization is dubious. That is why we do not include
the loop contributions in this present work, but will make a
detailed discussion in our coming paper.

Actually, even including the penguin contribution, the
theoretically estimated branching ratio of $BR(\Lambda_b\to
p+K^-)$ is still below the data and obviously smaller than that of
$BR(\Lambda_b\to p+\pi^-)$. If this measurement is valid and
approved by further experiments, it would be a new anomaly which
may hint an unknown mechanism which dominates the transition or
new physics beyond the standard model\footnote{We thank Dr. D.
Tonelli for bringing our attention to the new measurements of the
CDF collaboration on $BR(\Lambda_b\to p+\pi^-)$ and
$BR(\Lambda_b\to p+K^-)$. } and it is also consistent with the
result of \cite{Wangym}.

The good agreement of our results on the semileptonic decays of
$\Lambda_b$ to light baryon and several non-leptonic decay modes
with data indicates the following points.

First, the diqaurk picture: as we know, two quarks in a
color-anti-triplet attract each other and constitute a
Cooper-pair-like subject of spin 0 or 1. However, until now, many
theorists still doubt the justifiability of the diquark picture. It
is true that even though the diquark structure was raised almost as
early as the birth of the quark model, its validity or reasonability
of application is still in sharp dispute. In fact, it should be
rigorously testified by experiments. We have argued that for some
processes, the diquark picture may be more applicable than in the
others. Actually, in our case, we can convince ourselves that the
picture should apply. As aforementioned, diquark is only a spectator
in the transitions which we concern in this work, therefore its
inner structure may not affect the numerical results much. Secondly,
the produced baryon is very relativistic, i.e. very close to the
light-cone, generally the details of the inner structure of the
spectator diquark may not be important, this interpretation is
somehow similar to the parton picture which was conceived out by
Feynman and Bjorken long time ago. Namely at very high energy
collisions, the interaction among partons can be ignored at the
leading order, thus in our case the interaction between the quark
which undergoes a transition, and the spectator diquark should be
weak and negligible. Third is that the small effects caused by the
inner structure of the diquark may be partly included in the
parameter $\beta$ of the light-front wavefuction. The agreement with
data indicates that the diquark picture and the light-front quark
model indeed apply in the analysis of the heavy baryon transiting
into a light one.

Moreover, since we employ the factorization scheme to deal with the
non-leptonic decays, we find that to some modes, it works well, but
to some modes where loop contributions may dominate or just are
comparable to the tree contributions, the scenario encounters
serious challenges\cite{Wei}.

We further investigate the measurable polarization asymmetries.
Because the information on the polarization asymmetries may be more
sensitive to the model adopted in the theoretical calculations than
the decay width,  accurate measurements would discriminate various
models and indicate how to improve the details of the models.

Moreover, we also predict the  rates and asymmetries of several
similar modes of $\Lambda_b$ non-leptonic decays in the same model,
and the results are listed in Table IIIV of last section. The
numbers will be tested in the future.

Fortunately, the high luminosity at LHCb can provide large database
on $\Lambda_b$ and moreover, with great improvements of experimental
facility and detection technique, we expect that more and more
accurate measurements will be carried out in the near future and
theorists will be able to further testify,  improve, or even negate
our present models. Indeed, the baryons are much more complicated
than mesons, but careful studies on the processes where baryons are
involved would be very beneficial for getting better insight into
the hadron structure and underlying principles, especially the
non-perturbative QCD effects including the factorization and
plausibility of the diquark picture. The LHCb will be an ideal place
to do the job.

\section*{Acknowledgements:}

The work is supported by the National Natural Science Foundation of
China and the Special Grant for the PH.D program of the Chinese
Education Ministry, one of us (Ke) would like to thank the Tianjin
University for financial support.

\end{document}